\documentclass[twocolumn,english,aps,prb,showpacs,superscriptaddress,floats,amsmath,amssymb,floatfix]{revtex4}
\usepackage[T1]{fontenc}
\usepackage[latin9]{inputenc}
\usepackage{array}
\usepackage{graphicx}
\usepackage{amssymb}
\providecommand{\tabularnewline}{\\}
\usepackage{babel}
\usepackage[usenames]{color}
\allowdisplaybreaks
\begin{document}

\title{Series Expansion Analysis of a Frustrated Four-Spin-Tube}

\author{Marcelo Arlego}
\affiliation{Departamento de F\'isica, Universidad Nacional de La Plata,
C.C. 67, 1900 La Plata, Argentina}

\author{Wolfram Brenig}
\email{w.brenig@tu-bs.de}
\affiliation{Institut f\"ur Theoretische Physik,
Technische Universit\"at Braunschweig, 38106 Braunschweig, Germany }

\begin{abstract}
We study the magnetism of a frustrated four-leg spin-$1/2$ ladder with
transverse periodic boundary conditions: the frustrated four-spin tube
(FFST). Using a combination of series expansion (SE), based on the continuous
unitary transformation method and density-matrix renormalization group (DMRG) we
analyze the ground-state, the one-, and the two-particle excitations in the
regime of strong rung-coupling. We find several marked differences of the FFST
with respect to standard two-leg ladders. First we show that frustration
destabilizes the spin-gap phase of the FFST which is adiabatically connected to
the limit of decoupled rung singlets, leading to a first order quantum phase
transition at finite inter-rung coupling. Second, we show that apart from the
well-know triplon branch of spin-ladders, the FFST sustains additional
elementary excitations, including a singlon, and additional triplons. Finally we
find, that in the two-particle sector the FFST exhibits collective (anti)bound
states similar to two-leg ladders, however with a different ordering of the
spin-quantum numbers. We show that frustration has significant impact on the FFST
leading to a flattening of the ground-state energy landscape, a mass-enhancement
of the excitations, and to a relative enhancement of the (anti)binding
strength. Where possible we use DMRG to benchmark the findings from our
SE calculations, showing excellent agreement.
\end{abstract}


\maketitle

\section{Introduction}

Ever since it has been realized that there are 'surprises' on the
way from one- to two-dimensional quantum magnets \cite{Dagotto1996a},
spin ladders have attracted an enormous interest. Two-leg ladders,
such as Sr$_{14-x}$Ca$_{x}$Cu$_{24}$O$_{41}$ \cite{Uehara1996a},
have been under intense scrutiny mainly because of the potential interplay
between their spin-gaped ground state and the occurrence of superconductivity
\cite{Dagotto1999a,Johnston2000b}. N-leg ladders with $N>2$ are of
particular relevance, not only because they allow for a generalized
test of Haldane's conjecture \cite{Haldane1983a}, but also because
new, \emph{tube-like} lattice structures can be realized if periodic
transverse boundary conditions apply, such as in
{[}(CuCl$_{2}$tachH)$_{3}$Cl]Cl$_{2}$
\cite{Schnack2004a} and CsCrF$_{4}$ \cite{Manaka2009a} with $N=3$,
Cu$_{2}$Cl$_{4}$$\cdot$D$_{8}$C$_{4}$SO$_{2}$ \cite{Garlea2008a,Zheludev2008a}
with $N=4$, and Na$_{2}$V$_{3}$O$_{7}$ \cite{Millet1999a} with
$N=9$. For $N=3$, magnetic frustration surfaces as an additional
ingredient, already for tubes with only nearest neighbor exchange,
leading to a rich variety of phenomena not present in two-leg ladders
\cite{Kawano1997a,Sakai2000a,Fouet2006a,Nishimoto2008a}.

\begin{figure}[tb]
\begin{centering}
\includegraphics[width=0.7\columnwidth]{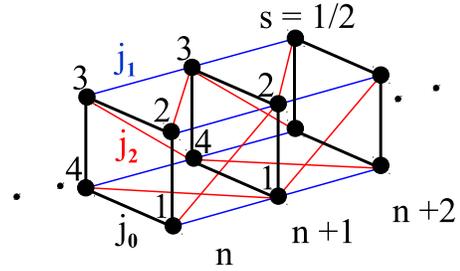}
\par\end{centering}
\caption{\label{fig1}
Frustrated four-spin tube. Solid circles represent spin-$1/2$
moments. Plaquettes (bold black lines) are coupled by nearest ($j_{1}$) and next
nearest ($j_{2}$) antiferromagnetic exchange, blue and red lines. On-plaquette
couplings ($j_{0}$) are set to unity.}
\end{figure}

Experimentally, the four-spin tube Cu$_{2}$Cl$_{4}$$\cdot$D$_{8}$C$_{4}$SO$_{2}$ has
been suggested to display frustrating antiferromagnetic next-nearest neighbor
exchange \cite{Garlea2008a}. Theoretically, {\em unfrustrated} four-spin tubes
have been considered in two seminal papers \cite{Cabra1998a,Kim1999}, both, in
the weak and the strong rung-coupling limit.  Magnetic frustration, however, has not
been considered in these studies. Therefore, in this work, we perform a first
analysis of a frustrated four-spin tube (FFST)
\begin{equation}
H=\sum_{lm}j_{lm}\mathbf{S}_{l}\cdot\mathbf{S}_{m}\,,
\label{eq:1}\end{equation}
with a lattice structure and exchange couplings $j_{lm}$ as shown in
Fig.\ref{fig1}. Spin-$1/2$ moments are located on the solid circles and all
couplings, $j_{0}$, $j_{1}$ and $j_{2}$ are antiferromagnetic (we set
$j_{0}=1$ hereafter). This FFST is simpler than the one proposed for
Cu$_{2}$Cl$_{4}$$\cdot$D$_{8}$C$_{4}$SO$_{2}$, where only part of the surface
squares experience diagonal exchange, and the leg-couplings seem to be
in-equivalent \cite{Garlea2008a}. Apart its relation to existing
materials, the FFST is of interest as a 1D variant of the anisotropic triangular
lattice on the torus with four site circumference, i.e. cutting the tube
longitudinally one obtains an anisotropic triangular lattice strip.

For $j_{1,2}\ll1$, the FFST resembles a chain of weakly coupled four-spin
plaquettes each of which displays a singly degenerate singlet ground
state, separated by a gap of $j_{0}=1$ from the first excited triplet. Therefore
perturbation theory in $j_{1,2}$ applies. Motivated by this
we investigate the FFST by series expansion (SE) in $j_{1,2}$.
Moreover, we corroborate our approach and gauge our SE results
by employing density-matrix renormalization group (DMRG) calculations.
The structure of the paper is as follows. In section \ref{II} we
clarify the region of applicability of the SE. Section \ref{III}
details our SE method. Results are presented in section \ref{IV},
including the ground state energy of the FFST in sub-section \ref{IVa},
the one-particle excitations in sub-section \ref{IVb}, as well as
two-particle states in \ref{IVc}. Conclusions are presented in
section \ref{V}. For completeness, technicalities of the two-particle
SE calculations are included in the appendix \ref{app1}.

\section{Coupled Plaquette Regime\label{II}}

Proper application of SE hinges on the adiabatic renormalization of
the bare starting state in terms of the coupling constants. In case
of intervening second order quantum phase transitions, the SE can
be used directly to limit its range of applicability in terms of diverging
susceptibilities or vanishing elementary excitation gaps. In case
of a discontinuous or first order transitions, SE based on a single
bare ground state fails to signal any transition. To put our SE
on firm grounds \emph{a-priori}, we therefore search for potential
first-order quantum phase-transitions of the FFST at small $j_{1,2}$.

\begin{figure}[t]
\begin{centering}
\includegraphics[width=.9\columnwidth]{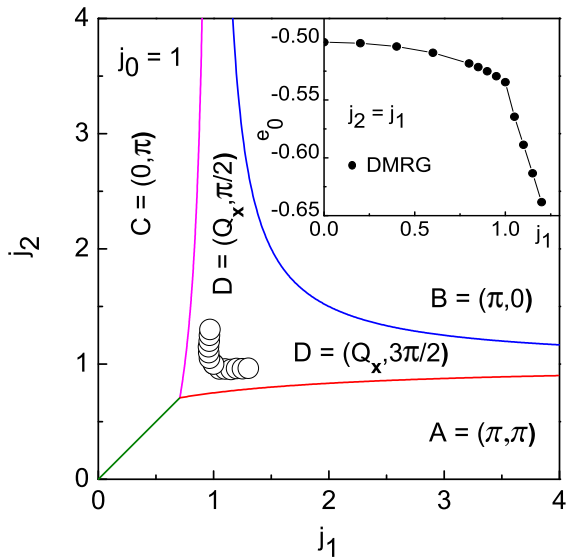}
\par\end{centering}
\caption{\label{fig2}
Phase diagram of the FFST. Solid lines: transitions at the {\em classical} level
(all are discontinuous). Wave vectors A, B, C and D label pitch angle of classical
phase.  Circles: first-order {\em quantum} critical line from DMRG (see text).
Inset: ground state energy versus $j_1$ from DMRG for the quantum model at
$j_2=j_1$, showing first order transition at $j_{1}=j_{1}^{c}\approx 1$.}
\end{figure}

To this end, it is instructive to first consider the classical phase diagram of
the FFST. We allow the spin structure to be a spiral, which, due to $SU(2)$
symmetry can be considered to be planar
$\mathbf{S}=S(\cos(\mathbf{Q}\cdot\mathbf{r}),\sin(\mathbf{Q}\cdot\mathbf{r}),0)$
with $\mathbf{r}=l_{x}\mathbf{R}_{x}+l_{y}\mathbf{R}_{y}$, where
$\mathbf{R}_{x,y}=(1,0),(0,1)$, $l_{x}\in\mathbb{N}$, and $l_{y}=[1,\ldots4]$.
The transverse pitch vector $Q_{y}$ has to be discretized according to
$(0,1,2,3)\pi/2$. The ground state energy is
$e_{g}^{c}=\cos(Q_{y})+j_{1}\cos(Q_{x})+j_{2}\cos(Q_{x}+Q_{y})$.  From this,
four classical phases result, shown in Fig. \ref{fig2}: a $(\pi,\pi)$
antiferromagnet for $j_{2}\leqslant(1+2j_{1})/(2(j_{1}+1)),\,\wedge\,
j_{2}\leqslant j_{1}$ (region A), a $(0,\pi)$ columnar antiferromagnet for
$j_{2}\geqslant(1-2j_{1})/(2(j_{1}-1)),\,\wedge\, j_{2}\geqslant j_{1}$ and
$j_{1}<1$ (region C), a $(\pi,0)$ columnar antiferromagnet for
$j_{2}\geqslant(2j_{1}-1)/(2(j_{1}-1)),\,\wedge\, j_{1}>1$ (region B). In the
remaining region D, the energy is minimized by two degenerate incommensurate
spirals with pitch $(Q_{x},Q_{y})=(2\arctan(\alpha),\pi/2)$ and
$(2\pi-2\arctan(\alpha),3\pi/2)$, and
$\alpha=(j_{1}+\sqrt{j_{1}^{2}+j_{2}^{2}})/j_{2}$.  Due to $Q_{y}$'s
discretization all of the classical transitions are discontinuous.

While the quantum analog of this rather rich phase diagram clearly deserves
future analysis, our SE is confined to the region of
$j_{1,2}\lesssim1$. Therefore we focus only this region regarding potential
first order quantum phase transitions. First, the classical 'diagonal'
transition from the $(0,\pi)$ to the $(\pi,\pi)$ state has no quantum analog,
since it is connected to the region of $j_{1,2}\ll1$.  There, and instead of the
classical $(\pi,\pi)$ or $(0,\pi)$ antiferromagnets, the quantum model
shows a phase of weakly coupled plaquette-singlets and the quantum ground state
is protected by the singlet-triplet gap of the bare plaquette. To assess the
relevance to the quantum case of the classical transitions from the $(0,\pi)$
and $(\pi,\pi)$ states into the doubly-degenerate $(Q_x,\pi/2),$ $(Q_x,3\pi/2)$
phase, we resort to a DMRG analysis of the ground state energy. For this
we use the ALPS package \cite{alps}. Since we are only concerned with first
order transitions, we refrain from any detailed finite size scaling
analysis. We find that for $j_{1,2}\lesssim2$ there is negligible finite size
dependence of the ground state energy for FFSTs of lengths $L \approx 30\ldots40$,
i.e. $120\ldots160$ spins, and that $m=300\ldots400$ states kept in the density
matrix lead to $4\ldots5$ digits of precision, which is sufficient for our
purpose.

The inset in Fig. \ref{fig2} shows a typical result for the ground state energy
$e_0$ obtained from DMRG versus $j_1$ along the diagonal $j_1=j_2$.  Obviously
the ground state energy displays a kink at $j_{1}=j_{1}^{c}\approx 1$, which we
identify with a first order transition. The circles in Fig.  \ref{fig2}
summarize a scan of locations of this transition which we have performed. The
size of the circles is a rough measure of the numerical accuracy for the locations
of the transition. These locations are remarkably close to those of the
classical model. Below these transitions SE based on decoupled plaquettes is
applicable. Beyond $j_{1,2}\sim1.5$ the critical points are increasingly hard to
detect accurately from the DMRG data. We speculate that below the
first-order transition lines at $j_{1}^{c}(j_{2})$ and $j_{2}^{c}(j_{1})$ the
bare plaquette state is adiabatically connected to a Luttinger liquid for
$j_{2}\rightarrow\infty$ and $j_{1}\rightarrow\infty$, respectively.

\section{Series Expansion Method\label{III}}

The main focus of this work is on SE in terms of $j_{1}$
and $j_{2}$, starting from the limit of isolated plaquettes. To this
end we decompose the Hamiltonian of the FFST into\begin{eqnarray}
H & = & H_{0}+V\,,\nonumber \\
H_{0} & = & \sum_{n}h_{0,n}\,,\hphantom{aaa}V=V_{1}+V_{2}\,,\label{eq:2}\end{eqnarray}
where $h_{0,n}$ is the plaquette Hamiltonian at site $n$\begin{eqnarray}
h_{0,n} & = & [\mathbf{S}_{1}\cdot\mathbf{S}_{2}+\mathbf{S}_{2}\cdot\mathbf{S}_{3}+\mathbf{S}_{3}\cdot\mathbf{S}_{4}+\mathbf{S}_{4}\cdot\mathbf{S}_{1}]_{n}\label{eq:3}\end{eqnarray}
and the perturbation $V=V_{1}+V_{2}$ is given by\begin{eqnarray}
V_{1} & = & j_{1}\sum_{n}\sum_{i=1}^{4}\mathbf{S}_{i,n}\cdot\mathbf{S}_{i,n+1},\nonumber \\
V_{2} & = & j_{2}\sum_{n}(\mathbf{S}_{1,n}\cdot\mathbf{S}_{2,n+1}+\mathbf{S}_{2,n}\cdot\mathbf{S}_{3,n+1}\nonumber \\
 &  & +\mathbf{S}_{3,n}\cdot\mathbf{S}_{4,n+1}+\mathbf{S}_{4,n}\cdot\mathbf{S}_{1,n+1})\,.\label{eq:4}\end{eqnarray}
The eigenstates of an isolated plaquette consist of four equidistant
energy levels $E_{n}=q_{n}-2$, labeled by the quantum number $q_{n}=0,..,3$,
and can be classified according to the total and the $z$-component
of the plaquette spin $\mathbf{\mathbf{S}}_{n}=\sum_{i=1}^{4}\mathbf{S}_{i,n}$.
Table \ref{tab1} lists that the ground state is a singlet, the first
excited state at $q_{n}=1$ is a triplet, the $q_{n}=2$ sector is
composed of a singlet and two triplets, and for $q_{n}=3$ one quintet
remains.

\begin{table}[tb]
\begin{centering}
\begin{tabular}{|l|c|c|c|c|c|}
\hline
State  & $q_{n}$  & $E_{n}$ & $S$  & $S_{z}$ & Idx \tabularnewline
\hline
$|s_{0}\rangle$  & 0  & -2 & 0  & 0 & 0 \tabularnewline
\hline
$|t_{0}^{S_{z}}\rangle$  & 1  & -1 & 1  & -1,0,1 & 1,2,3 \tabularnewline
\hline
$|s_{1}\rangle$  & 2  & 0 & 0  & 0 & 4 \tabularnewline
\hline
$|t_{1}^{S_{z}}\rangle$  & 2  & 0 & 1  & -1,0,1 & 5,6,7 \tabularnewline
\hline
$|t_{2}^{S_{z}}\rangle$  & 2  & 0 & 1  & -1,0,1 & 8,9,10 \tabularnewline
\hline
$|q^{S_{z}}\rangle$  & 3  & 1 & 2  & -2,...,2 & 11,...,15 \tabularnewline
\hline
\end{tabular}
\par\end{centering}
\caption{\label{tab1}{
Spectrum of the single plaquette. It consists of four equidistant energy levels
$E_{n}=q_{n}-2$ labeled by the quantum number $q_{n}=0,..,3$, the total spin
$S$, and its $z$-component $S_z$.  The last column enumerates the states.}}
\end{table}

In turn, $H_{0}$ displays an equidistant spectrum, labeled by $Q=\sum_{n}q_{n}$.
At $V=0$, the $Q=0$ sector refers to the \textit{\emph{unperturbed}}
singlet ground state $|0\rangle=\otimes_{n}|s_{0,n}\rangle$ composed
of $q_{n}=0$ singlets on all plaquettes. The $Q=1$ sector comprises
single $|t_{0,m}^{S_{z}}\rangle$ triplets inserted into $|0\rangle$
at site $m$. The $Q\geq2$ sectors are of multiparticle nature. The
perturbation $V$ can be rewritten as $V=\sum_{i=1}^{2}j_{i}\sum_{n=-N}^{N}T_{n}^{i}$,
where $T_{n}$ represent raising ($n>0$) or lowering ($n<0$) operators
within the spectrum of $H_{0}$. For the FFST we find $N=4$.

It has been shown quite generally \cite{Knetter2000a}, that models
with the preceding type of spectrum allow for SE through a continuous
unitary transformation (CUT) using the flow equation method of Wegner
\cite{Wegner1994a}. The basic idea is to transform $H$ onto an effective
Hamiltonian $H_{eff}$ which is \emph{block-diagonal} in the quantum
number $Q$. This transformation can be achieved exactly order-by-order
in $j_{1,2}$ leading to\begin{equation}
H_{\mathrm{eff}}=H_{0}+\sum_{n,0\leq m\leq n}C_{n,m}j_{1}^{n-m}j_{2}^{m}\,,\label{eq:5}\end{equation}
where the $C_{n,m}$ are weighted products of the $T_{n}^{i}$ operators
which conserve the $Q$-number and have their weights determined by
recursive differential equations, see \cite{Knetter2000a} for details.
Due to $Q$-number conservation several observables can be accessed
directly from $H_{eff}$ in terms of a SE in $j_{1,2}$. For systems
with coupled spin-plaquette CUT SE has been used for one \cite{Arlego2006a},
two \cite{Arlego2008a,Arlego2007a,Brenig2004a,Brenig2002aa} and three
\cite{Brenig2003a} dimensions.

\section{Results\label{IV}}

In this Section we present our findings from SE up to $Q=2$, sectioning the
discussion according to A. the ground-state energy, B. the one-, and C. the
two-particle excitations. To assess the quality of the SE we complement our
analysis by DMRG calculations for selected cases.

\subsection{Ground State Energy\label{IVa}}
First we consider the ground state energy $E_{0}$.
$Q$-conservation leads to \begin{equation} E_{0}=\langle0|H_{{\rm
eff}}|0\rangle\,,\label{eq:6}\end{equation} where $|0\rangle$ is the
\emph{unperturbed} ground state.

\begin{figure}[tb]
\begin{centering}
\includegraphics[width=0.7\columnwidth]{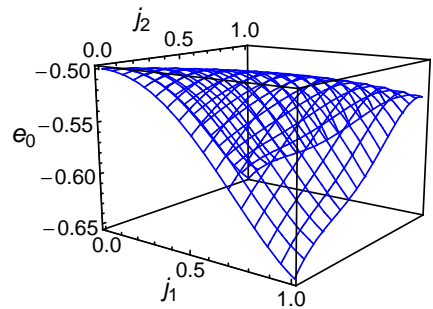}
\par\end{centering}
\caption{\label{fig3}
Ground state energy per site $e_{0}$ versus $j_{1}$ and $j_{2}$, showing a
monotonously decreasing behavior in the parameter range shown.  Along the
line of maximum frustration $j_{1}=j_{2}$, the energy gain is smallest.}
\end{figure}

Evaluating this matrix element on chains with periodic boundary conditions (PBC)
of a length $L$, sufficient not to allow for wrap-around of graphs
with length $N$, i.e. $L=N+1$, one can obtain \emph{analytic }SEs
which are valid to $O(N)$ with respect to eqn. (\ref{eq:5}) in the
thermodynamic limit. Evaluating the ground state energy per
spin $e_{0}=E_{0}/(4L)$ up to $O(7)$ we get
\begin{eqnarray}
\lefteqn{e_{0}=-\frac{1}{2}-\frac{17j_{1}^{2}}{96}-\frac{17j_{2}^{2}}{96}+\frac{j_{1}j_{2}}{3}-\frac{25j_{1}^{3}}{384}-\frac{25j_{2}^{3}}{384}+\frac{j_{1}^{2}j_{2}}{16}}\nonumber \\
 &  & +\frac{j_{1}j_{2}^{2}}{16}+\frac{1919j_{1}^{4}}{96768}+\frac{1919j_{2}^{4}}{96768}-\frac{313j_{1}^{3}j_{2}}{1728}-\frac{313j_{1}j_{2}^{3}}{1728}\nonumber \\
 &  & +\frac{15457j_{1}^{2}j_{2}^{2}}{48384}+\frac{1510499j_{1}^{5}}{27095040}+\frac{1510499j_{2}^{5}}{27095040}-\frac{39353j_{1}^{4}j_{2}}{188160}\nonumber \\
 &  & -\frac{39353j_{1}j_{2}^{4}}{188160}+\frac{4129273j_{1}^{3}j_{2}^{2}}{27095040}+\frac{4129273j_{1}^{2}j_{2}^{3}}{27095040}\nonumber \\
 &  & +\frac{522374480359j_{1}^{6}}{14748372172800}+\frac{522374480359j_{2}^{6}}{14748372172800}-\frac{92947333j_{1}^{5}j_{2}}{4267468800}\nonumber \\
 &  & -\frac{92947333j_{1}j_{2}^{5}}{4267468800}-\frac{1494532466633j_{1}^{4}j_{2}^{2}}{4916124057600}\nonumber \\
 &  & -\frac{1494532466633j_{1}^{2}j_{2}^{4}}{4916124057600}+\frac{9888732599j_{1}^{3}j_{2}^{3}}{17069875200}\nonumber \\
 &  & -\frac{535161937582507j_{1}^{7}}{24777265250304000}-\frac{535161937582507j_{2}^{7}}{24777265250304000}\nonumber \\
 &  & +\frac{3149204376698497j_{1}^{6}j_{2}}{12388632625152000}+\frac{3149204376698497j_{1}j_{2}^{6}}{12388632625152000}\nonumber \\
 &  & -\frac{274272578154571j_{1}^{5}j_{2}^{2}}{412954420838400}-\frac{274272578154571j_{1}^{2}j_{2}^{5}}{412954420838400}\nonumber \\
 &  &
 +\frac{1186859862395537j_{1}^{4}j_{2}^{3}}{2753029472256000}+\frac{1186859862395537j_{1}^{3}j_{2}^{4}}{2753029472256000}
\label{eq:7}\end{eqnarray}
Here, the first term corresponds to the bare energy per spin listed in table
\ref{tab1}. Since the FFST can be mapped onto an identical FFST with
$j_{1}\leftrightarrow j_{2}$ by a $\pi/2$-twist of the plaquettes around the
tube, one expects that $e_{0}(j_{1},j_{2})=e_{0}(j_{2},j_{1})$, which is
obviously fulfilled. Figure \ref{fig3} shows $e_{0}(0\leq j_{1}\leq1,0\leq
j_{2}\leq1)$ to be monotonously decreasing with $j_{1,2}$. Along the line
$j_{1}=j_{2}$, the energy gain is smallest. Speaking differently, along its
maximally frustrated direction in parameter space the energy landscape is
flattest.

\begin{figure}[tb]
\begin{centering}
\includegraphics[width=0.9\columnwidth]{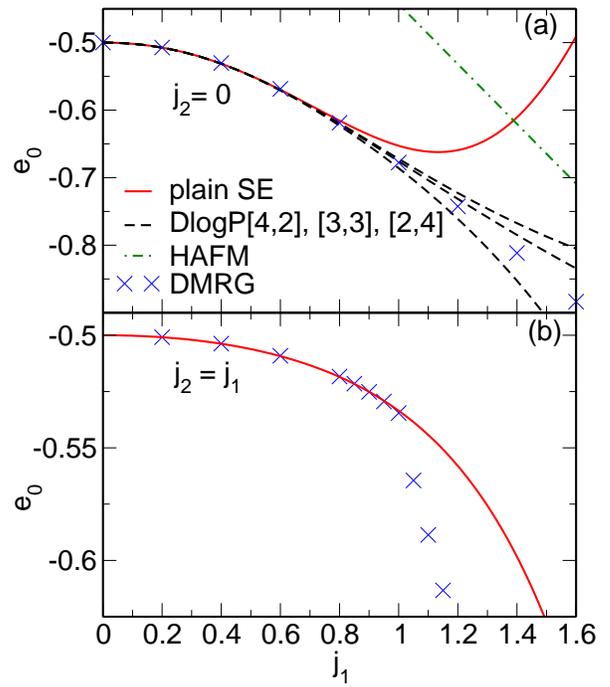}
\par\end{centering}
\caption{\label{fig4}
(a) Comparison between $e_{0}$ from plain SE (red solid lines) with DMRG for
$L=30$ and $m=300$ (blue crosses), along $j_{2}=0$.  Black dashed lines refer to
three DlogPad\'e approximants to eqn. (\ref{eq:7}). For reference the ground
state energy of the 1D HAFM is also depicted (green dash-dotted).  (b)
Comparison of plain SE (red solid lines) with DMRG (blue crosses) along
$j_{1}=j_{2}$. Up to the first-order transition discussed in section \ref{II},
the agreement is excellent.}
\end{figure}

In Fig. \ref{fig4} we asses the accuracy of eqn. (\ref{eq:7}) in several
ways. First, in panel (a) we compare $e_{0}$ with DMRG calculations along
$j_{2}=0$ where we expect no first-order transition. The DMRG results have been
obtained for $L=30$ and $m=300$. Obviously the agreement between SE and DMRG is
very good for $j_{1}\lesssim0.75$. In panel (b) we display a similar comparison
along $j_{1}=j_{2}$. Here the SE agrees very well with DMRG to even larger
values of $j_{1,2}$, however only up to the first-order transition discussed in
section \ref{II}. Third, in panel (a) we also include three DlogPad\'e
approximants to eqn. (\ref{eq:7}), which, similar to the DMRG, start to depart
from the SE for $j_{1}\gtrsim0.75$ and may be used to increase the interval of
confidence of the SE only slightly by $\sim10\%$ as evident from the DMRG
results. For reference panel (a) also depicts the ground state energy per site
of the 1D HAFM with exchange constant $j_{1}$ which is
$(-\ln(2)+1/4)j_{1}$. Evidently, in the parameter range we consider, the FFST is
far away from the decoupled chain limit.

\subsection{One-particle Excitations\label{IVb}}

Now we turn to the dispersion of one-particle eigenstates. By the latter we mean
excited eigenstates of the effective Hamiltonian which, apart from having fixed
$Q$, lattice momentum $k$, and total spin quantum numbers $S,m$, are linear
combinations of \emph{single} entries of table \ref{tab1} only. $Q=1$
eigenstates are single particle states by construction. For $Q\geqslant1$,
$Q$-conservation does not protect single states in table \ref{tab1} from decay
into \emph{two} states by virtue of $H_{eff}$. Eg. a $|s_{1l}\rangle$ singlet
with $Q=2$ at site $l$ could decay into two $|t_{0i(j)}^{S_{z}}\rangle$ triplets
with total spin $S=0$ and $Q=1+1$ at sites $i(j)$ (see also appendix
\ref{app1}). For the FFST however, and for $Q=2$, $S=0,1$ and up to $O(7)$, we
find that all matrix elements of $H_{eff}$ inducing such decay \emph{vanish}
identically.  This feature can be traced back to the $C_4$ symmetry of the
tube. In fact, we obtain that changing e.g. the exchange couplings $j_1$ into
$j_1'\neq j_1$ on one of the legs, renders the one-particle $Q=2$ states
unstable against decay. In summary, for each momentum $k$ the $Q=2$ spectrum
contains three genuine one-particle levels with $S=0,1$, all of which are
degenerate in $m$. For the remaining {\em two}-particle states with $Q=2$ we
refer to the next section.

For the rest of this section we label the one-particle states by
$|i\rangle^{Q,S_{n}}$, where $i$ refers to the plaquette's site and $S_{n}$ is
the total spin, where the index $n$ is only due to the fact that for $Q=2$ there
are two $S=1$ states, say, $n=a,b$. Spin-z quantum numbers $m$ are discarded
because of $SU(2)$ invariance. Due to $Q$-conservation and $SU(2)$ invariance
the effect of $H_{{\rm eff}}$ on $|i\rangle^{Q,S_{n}}$ is limited to
\begin{equation}
H_{{\rm
eff}}|j\rangle^{Q,S_{n}}=\sum_{i,S_{n}=S_{m}}t_{i}^{Q,S_{n},S_{m}}|j+i\rangle^{Q,S_{m}}\,,
\label{eq:8}
\end{equation}
which implies a shift in real space, and potentially a mixing of states of equal
$S$ with identical $Q$. The hopping amplitudes $t_{i}^{Q,S_{n},S_{m}}$ do not
depend on $j$ due to translation invariance. Therefore, by Fourier
transformation $|k\rangle^{Q,S_{n}}=1/\sqrt{L}\sum_{j}\exp(-i\, k\,
j)|j\rangle^{Q,S_{n}}$ we get the dispersion from
\begin{eqnarray}
\lefteqn{E_{Q,S_{n},S_{m}}^{\textrm{1pt}}(k)={}^{Q,S_{n}}
\langle k|H_{\textrm{eff}}|k\rangle^{Q,S_{m}}-E_{0}\,\delta_{S_{n},S_{m}}}\nonumber \\
 &  &
\hphantom{aaaaaa}=\tilde{t}_{0}^{Q,S_{n},S_{m}}+2\sum_{i}t_{i}^{Q,S_{n},S_{m}}g(ik)\,,
\label{eq:9}\end{eqnarray}
where, obviously $t_{i}^{Q,S_{n},S_{m}}=t_{-i}^{Q,S_{n},S_{m}}$ for
$S_{n}=S_{m}$. However for $S_{n}\neq S_{m}$, i.e. for the two $Q=2$, $S=1$
states, we find $t_{i}^{Q,S_{n},S_{m}}=-t_{-i}^{Q,S_{n},S_{m}}$.  In turn,
$g(x)$ is $\cos[i\,\sin](x)$ for $S_{n}$-{[}off]diagonal transitions.

To obtain hopping amplitudes valid to $O(N)$, in the thermodynamic limit, the
$t_{i}^{Q,S_{n},S_{m}}$ and the ground state energy
$E_{0}=\langle0|H_{\mathrm{eff}}|0\rangle$ in eqn.(\ref{eq:9}) have to be
evaluated on clusters with open boundary conditions (OBC), large enough to
incorporate $N$-th order graphs for hopping processes of distance $i$
\cite{Knetter2000a}. Depending on $i$, these clusters are of either of length
$N$ or $N+1$. We have calculated \emph{analytic }expressions \cite{ElecAvail}
for $E_{Q,S_{n},S_{m}}^{\textrm{1pt}}(k)$ to $O(7)$ in $j_{1,2}$. For
$(Q,S)\neq(2,1)$ eqn. (\ref{eq:9}) is already diagonal in $S_{n}$, $S_{m}$ with
eigenvalues $E_{Q,S_{n}}^{\textrm{1pt}}(k)\equiv
E_{Q,S_{n},S_{n}}^{\textrm{1pt}}(k)$.  Only for $(Q,S)=(2,1)$ eqn. (\ref{eq:9})
displays a $2\times2$-matrix structure, referring to $n=a,b$, with eigenvalues
$E_{Q,1a}^{\textrm{1pt}}(k)$ and $E_{Q,1b}^{\textrm{1pt}}(k)$.

\begin{figure}[t]
\begin{centering}
\includegraphics[width=1\columnwidth]{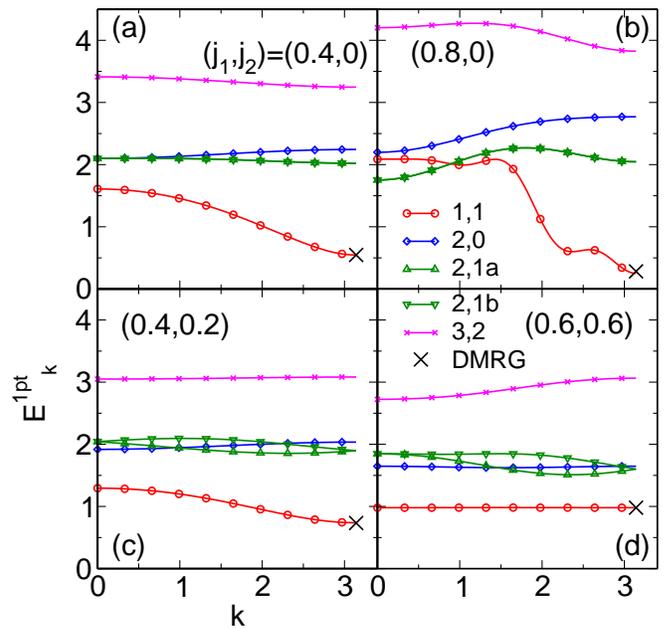}
\par\end{centering}
\caption{\label{fig5}Genuine one-particle dispersions on the FFST for $Q\leq2$
at various values of $(j_{1},j_{2})$. These comprise three triplets
(red circle, green triangle up, green triangle down) and one singlet
(blue diamond). The $Q$=3-quintet (magenta x) has not been tested
against multi-particle decay and is only shown for reference. DMRG
results at zone-boundaries are shown with large crosses.}
\end{figure}

In Fig. \ref{fig5} we show the one-particle dispersions for selected
values of $j_{1,2}$. This figure displays only bare SE results and
no Pad\'e extrapolations. Several comments are in order. First, the
figure does not only display $E_{Q,S_{n}}^{\textrm{1pt}}(k)$ for
all $S$ and $Q\leq2$, but for curiosity also the quintet with $Q=3$,
assuming that the latter does not decay into multi-particle states
- which we have not checked.

As for the ground state energy, since exchanging $j_{1}\leftrightarrow j_{2}$
maps the FFST onto an equivalent one by a $\pi/2$-twist there are related
symmetries of the one-particle dispersions. These respect the additional fact,
that the single-particle states on the bare plaquette are the eigenstates of a
four-site spin-$1/2$ chain with PBC, which carry one out of four momenta
$k_{n\perp}=n\pi/2$ with $n=0,1,2,3$ transverse to the FFST. In turn, exchanging
$j_{1}\leftrightarrow j_{2}$ maps the one-particle dispersions onto identical
ones up to a shift of the Brillouin zone by one out of $k_{n\perp}$, and may
also exchange the dispersion branches for the degenerate bare $Q,S=2,1$ states.
We have checked this to be fulfilled by all $E_{Q,S_{n}}^{\textrm{1pt}}(k)$.
Eg., for $(Q,S)=(1,1)$ we have $E_{Q,S_{n}}^{\textrm{1pt}} (k,j_{1},j_{2}) =
E_{Q,S_{n}}^{\textrm{1pt}} (k+\pi,j_{2},j_{1})$ and for $(Q,S)=(2,1a[b])$ we
have $E_{Q,1a[b]}^{\textrm{1pt}} (k,j_{1},j_{2}) = E_{Q,1b[a]}^{\textrm{1pt}}
(k+[-]\pi/2,j_{2},j_{1})$.  Additionally $E_{Q,1a[b]}^{\textrm{1pt}}
(k,j_{1},j_{2})$ are degenerate at $j_{2}=0$.

Considering the elementary triplet dispersion $E_{1,1}^{\textrm{1pt}}(k)$ at
$j_{2}=0$, we observe a zone-boundary gap in Fig. \ref{fig5} (a) which decreases
as $j_{1}$ increases, see panel (b).  For four-leg ladders without frustration,
i.e. $j_{2}=0$, the analytic expression for the gap we get is
\begin{eqnarray}
\lefteqn{E_{1,1}^{\textrm{1pt}}(\pi,j_{1},0)=1-\frac{4}{3}j_{1}+
\frac{41}{108}j_{1}^{2}+\frac{349}{1296}j_{1}^{3}}\nonumber \\
 & \hphantom{aaaaaaa} & +\frac{4596401}{39191040}j_{1}^{4}-
\frac{169497997}{4702924800}j_{1}^{5}\hphantom{aaaaaaa}\nonumber \\
 &  & -\frac{689874137639377}{2986545364992000}j_{1}^{6}\nonumber \\
 &  & -\frac{8430165345498432721}{45156565918679040000}j_{1}^{7}\,,
\label{eq:91}\end{eqnarray}
which, in passing, improves earlier SE to $O(4)$ on \emph{unfrustrated }four-leg
ladder \cite{Cabra1998a} by three orders. The dispersion
$E_{1,1}^{\textrm{1pt}}(k)$ shows rather strong oscillations in panel (b). These
are not related to convergence issues of the SE, but are robust features of
hopping to more than only nearest-neighbors.

Increasing $j_{2}$ from zero, at fixed $j_{1}$, as from Fig. \ref{fig5} (a) to
(c) the bandwidth of $E_{1,1}^{\textrm{1pt}}(k)$ is reduced. On the line of
maximum frustration, i.e. at $j_{1}=j_{2}$ in panel (d), it is only very small,
albeit \emph{not} exactly zero, namely $O(j_{1}^{4})$.  A similar tendency can
be observed in all other $(Q,S)$ sectors, yet less pronounced.

To assess the quality of the SE, we have also calculated the elementary
$(Q,S)=(1,1)$ gap by DMRG. The results are shown by the large crosses
at the zone-boundaries in Fig. \ref{fig5}. As is obvious, the agreement
is very good in all four panels. This is particularly noteworthy for
panel (b), corroborating the preceding statement, that the oscillations
in $E_{1,1}^{\textrm{1pt}}(k)$ for larger values of $j_{2}$ are
robust features of the SE and no convergence issue.

\subsection{Q=2 Two Particle States\label{IVc}}

\begin{table}[tb]
\centering{}\begin{tabular}{|>{\centering}p{0.3in}|l|}
\hline
\noalign{\vskip\doublerulesep}
S & $|ij\rangle^{Sm}\,,\,|i\rangle^{Sm}$\tabularnewline[\doublerulesep]
\hline
\noalign{\vskip\doublerulesep}
\hline
2 & $\sum_{n}C_{1n,1m-n}^{2m}|t_{0i}^{n}t_{0j}^{m-n}\rangle$\tabularnewline
\hline
1 & $\sum_{n}C_{1n,1m-n}^{1m}|t_{0i}^{n}t_{0j}^{m-n}\rangle\,,\,|t_{1i}^{m}\rangle\,,\,|t_{2i}^{m}\rangle$\tabularnewline
\hline
0 & $\sum_{n}C_{1n,1m-n}^{00}|t_{0i}^{n}t_{0j}^{m-n}\rangle\,,\,|s_{1i}\rangle$\tabularnewline
\hline
\end{tabular}\caption{\label{tab2}$Q=2$ states from the bare plaquette. $C_{1n,1m-n}^{Sm}=\langle1n,1m-n|Sm\rangle$
are Clebsch-Gordan coefficients. }
\end{table}

In this section we focus on two-particle states generated from two one-particle
triplets out of the $(Q,S)=(1,1)$ sector. Apart from generating a continuum of
states at fixed total momentum with respect to the relative momentum,
interactions may lead to additional collective (\emph{anti})\emph{bound states
}which split off from the continuum.  The existence of such collective states is
know for simpler 1D quantum magnets, such as chains and two-leg ladders
\cite{Jurecka2000c,Trebst2000,Knetter2001a,Zheng2001a,Brenig2001z,Knetter2004b,Schmidt2004a}.
Here we show that similar states exist also on the FFST. To this end we evaluate
the two-particle spectrum following ideas developed for dimer-SE of two-leg
ladders \cite{Knetter2004b}. For the sake of completeness technical details
of this procedure are revisited in appendix \ref{app1}.

On the FFST, the $Q=2$ sector allows for one-particle \emph{and} two-particle
states. This is different from dimer systems where only two-particle states
occur for $Q=2$ \cite{Knetter2004b}. The real-space representations of all $Q=2$
states with proper spin quantum numbers $S,m$ are listed in table
\ref{tab2}. Here the genuine two-particle states are labeled by
$|i,j\rangle^{Sm}$, where $i,j$ refers to sites on the 1D lattice and $S,m$ to
total spin, and spin-$z$ quantum numbers. As discussed in the preceding section
the one- and two-particle excitations at $Q=2$ do not mix. Therefore we focus on
the genuine two-particle excitations hereafter.

Because of translational invariance the two-particle states can be
classified according to a center-of-mass momentum $q$ and a relative
distance $d$
\begin{equation}
|q,d\rangle^{Sm}=\frac{1}{\sqrt{L}}\sum_{r}e^{iq(r+d/2)}|r,r+d\rangle^{Sm}
\label{eq:10}\end{equation}
These states have an exchange parity $P=(-1)^{S}$ for $d\rightarrow-d$,
i.e. $|q,-d\rangle^{Sm}=P|q,d\rangle^{Sm}$.  The effective Hamiltonian,
eqn. (\ref{eq:5}), will mix states at different $d$. The two-particle spectrum
is obtained from the sum of the one- and two-particle irreducible Hamiltonians
(see\cite{Zheng2001a,Knetter2004b} and appendix \ref{app1}), labeled by $H_{1}$
and $H_{2}$, with \emph{analytic} matrix elements calculated by SE from the
states of eqn. (\ref{eq:10}). We have evaluated these matrix elements to
$O(6)$. For each $q$, this leads to an eigenvalue problem with a rank set by the
number of lattice sites (or relative momenta) $L$. This can be diagonalized
numerically on finite, but large lattices. See appendix \ref{app1} for
definitions and formal details.

\begin{figure}
\centering{}\includegraphics[width=0.75\columnwidth]{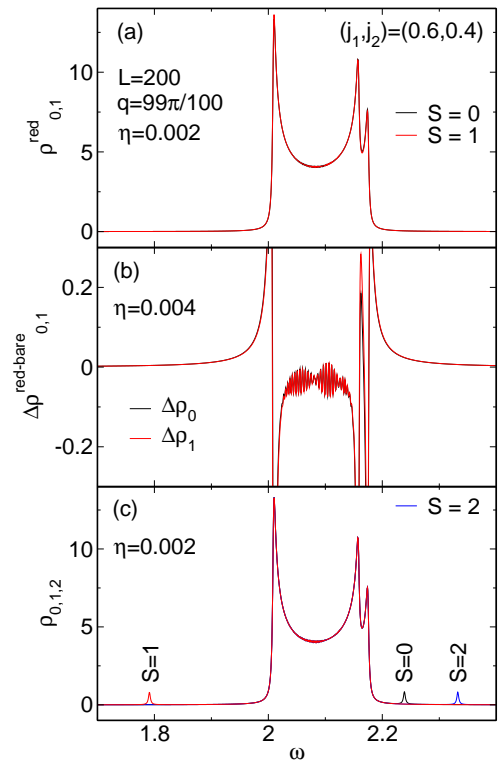}\caption{\label{fig6}Two-particle density of states close to the zone boundary
at $q=99\pi/100$ on an $L=200$ site FFST for $(j_{1},j_{2})=(0.6,0.4)$.
Delta functions at each two-particle energy broadened by $\eta$.
(a) \emph{reducible} spectrum for S=0 and S=1. (b) difference between
bare and \emph{reducible} spectrum for S=0 and S=1. (c) complete spectra
for S=0,1,2.}
\end{figure}

In Fig. \ref{fig6} we show the two-particle spectrum
$\rho_{S}(q,\omega)=\sum_{p}\delta(\omega-E_{S,m}^{2pt}(q,p))$ at fixed total
momentum $q$ and spin $S,m$, summed over the relative momentum $p$, where
$E_{S,m}^{2pt}(q,p)$ are the two-particle eigenenergies.  We choose
$(j_{1},j_{2})$ = $(0.6,0.4)$ and $q\simeq\pi$ as an example. To analyze the
effects of the interaction we consider $E_{S,m}^{2pt}(q,p)$ at various levels of
approximation. First, we set the irreducible two-particle interactions $H_{2}$
to zero - i.e. no 'actual' interactions occur. The corresponding spectrum is
shown in Fig. \ref{fig6} (a) for $S=0$ and $S=1$. These spectra are
indistinguishable on the scale of this plot. The van-Hove singularities are
related to the extrema of the one-particle dispersion in the $(Q,S)=(1,1)$
sector as in Fig. \ref{fig5}. Even though the irreducible one-particle
interaction $H_{1}$ will act only on one of the two particles, this does
\emph{not} imply, that the spectra in Fig. \ref{fig6} (a) are identical to the
convoluted \emph{bare }spectra of two \emph{single}
$|t_{0}^{S_{z}}\rangle$-particles at fixed total momentum $q$ and finite
$L$. This is due to the hard-core constraint which forbids double occupation of
plaquettes by $|t_{0}^{S_{z}}\rangle$-particles.  This constraint is encoded in
the matrix elements of $H_{1}$ when acting on the two particle sector. Since
there are $L$ double occupancies within $L^{2}$ two-particle states, their
removal amounts to a $1/L$ effect on the integrated spectral weight at fixed
total momentum $q$. At fixed total and relative momentum, this effect will be
\emph{quantitatively} identical for two-particle states of identical exchange
parity $(-)^{S}$.  For this reason Figs. \ref{fig6} (a) and (b) contain only
$S=0$ and 1 sates. Figure \ref{fig6} (b) depicts the difference between the
\emph{bare} spectra and that obtained from $H_{1}$ for $S=0$ and $S=1$ for
$L=200$. Clearly the integrated spectral weight for both values of $S$ is finite
but of $O(1/L)$ at most. Differences are visible between the $S=0$ and $S=1$
spectrum - albeit very small.  Figure \ref{fig6} (a) shows that \emph{no}
(anti-)bound state arises in the two-particle sector solely due to the hard-core
repulsion.

\begin{figure}[tb]
\begin{centering}
\includegraphics[width=0.85\columnwidth]{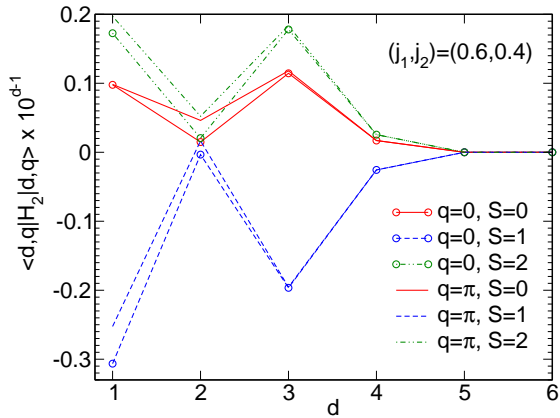}
\par\end{centering}
\caption{\label{fig7}Diagonal element of the irreducible
interaction $\langle q,d|H_{2}|q,d\rangle$ versus relative distance of two
particles for $S=0,1$, and 2, and for $q=0$ and $\pi$, multiplied by
$10^{d-1}$. (The matrix elements of $H_{2}$ at $O(6)$ are zero for $d=5,\,6$.)}
\end{figure}

Next we consider the complete $Q=2$ spectrum in Fig. \ref{fig6} (c) for $S=1,2$
and $3$, including the irreducible two-particle interaction $H_{2}$. The latter
is \emph{spin-dependent} and \emph{longer-ranged}.  This figure clearly
demonstrates a main point of this section, i.e.  the occurrence of three
split-off (\emph{anti})-\emph{bound} states (above) below the continuum
depending on their spin. Apart from that, and similar to the hard-core
constraint $H_{2}$ leads to an $O(1/L)$ redistribution within the two-particle
continuum. The total momentum $q$ in Fig. \ref{fig6} has been chosen, such as to
evidence the (anti)-bound states clearly. Depending on $q$ they can approach and
also merge with the continuum.

A qualitative understanding of binding versus anti-binding of the collective
two-particle states can be obtained from considering the diagonal matrix
elements $^{Sm}\left\langle q,d\right|H_{2}|q,d\rangle^{S^{\prime}m^{\prime}}$
of two particles versus their relative distance $d$ at fixed total momentum
$q$. This matrix element is $\propto\delta_{S^{\prime}S}\delta_{m^{\prime}m}$
and independent of $m$. It is shown in Fig. \ref{fig7} for $j_{1,2}$ identical
to that of Fig. \ref{fig6}. First, this figure demonstrates that the irreducible
two-body interactions are very rapidly decaying as a function of the relative
distance of the particles (note the scaling of the matrix elements by $10^{d-1}$
on the y-axis). Second, and for both total momenta depicted, the interactions
for $S=1$ are attractive, while those for $S=0,2$ are repulsive. This is
consistent with binding versus anti-binding of the collective states in Fig.
\ref{fig6}. Yet, one should keep in mind the matrix elements
$^{S^{\prime}m^{\prime}}\left\langle q,d^{\prime}\neq
d\right|H_{2}|q,d\rangle^{Sm}$ which are off-diagonal in $d^{\prime},d$. We have
not tested if these invalidate the simplified argument given here. Finally, we
emphasize, that the ordering of anti-bound versus bound states on the FFST is
different from that on plain two-leg ladders. In the latter case two-particle
singlets also form {\em bound} states with a binding energy larger than that of
the two-particle triplets.

\begin{figure}[t]
\begin{centering}
\includegraphics[width=1.0\columnwidth]{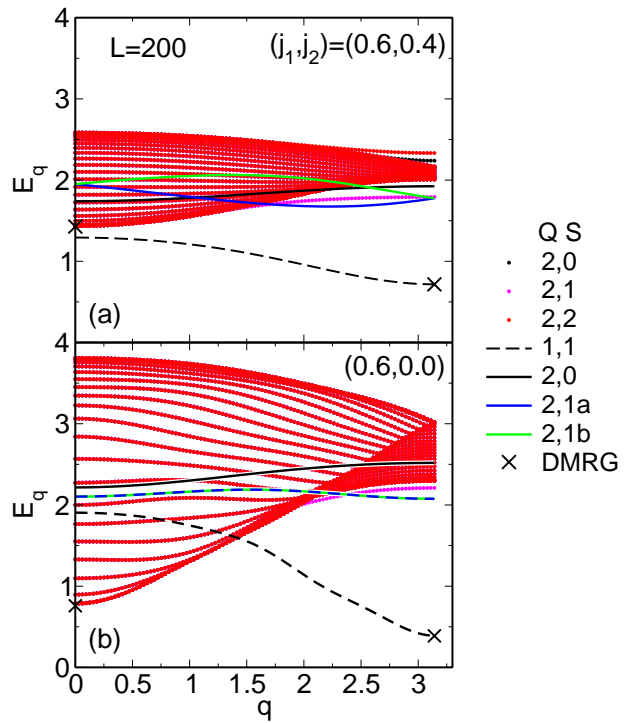}
\par\end{centering}
\caption{\label{fig8}Momentum resolved spectrum of all states on the FFST with $Q\leq2$ for
$(j_{1},j_{2})=(0.6,0.4)$ and $(0.6,0)$. Apart from four genuine one-particle
states, i.e. three triplets (black dashed, blue solid, green solid), and one
singlet (black solid), there are three two-particle continua with S=0,1,2, and
three collective two-particle states: a bound triplet (magenta dots), an
anti-bound singlet
(black dots) and an anti-bound quintet (red dots).  Only $\sim 10$\% of the
scattering states between the lower and upper bounds of the continua are
depicted to avoid graphic cluttering.  DMRG results for one and two-particle
states at zone-boundaries(centers) are depicted with large crosses.}
\end{figure}

In Fig. \ref{fig8} we show the dispersion of all one and two-particle states
from the $Q=1$ and $Q=2$ sector as a function of total momentum for two sets of
$j_{1,2}$, obtained by SE, together with DMRG results for one and two-particle
states at the zone-boundary(center).  This figure is another main result of our
paper and summarizes several aspects. First, the spectrum is very rich and
consists of various discrete states and three superimposed two-particle
continua. This should be contrasted against plain two-leg ladders, which show a
less involved low-energy spectrum \cite{Trebst2000,Knetter2004b}.  Some of the
discrete states are genuine one-particle states and some are collective
two-particle states. For the parameters depicted, the collective two-particle
states are clearly visible only close to the zone boundary.  The potential
existence of critical parameters $j_{1,2}^{c}$ or wave-vectors $q^{c}$ for which
the collective states merge with the continua is unclear at present. The $Q=2$ one-particle states are contained almost completely within
the spectral range of the continua and split off from the latter only in a range
of $q$ similar to the collective states. This situation is also very different
from two-leg spin-ladders in which the states which split off from the continuum
close to the zone boundary are collective (anti)-bound states only.  From an
experimental point of view this may pose a challenge on discriminating between
such genuine one-particle and collective two-particle states.  Finally, the
effects of frustration are clearly visible in going from Fig. \ref{fig8} a) to
b): the complete spectrum shows a tendency to localize. However, in contrast to
the one-particle dispersions the two-particle interactions are {\em not} reduced
approaching the line of maximum frustration, $j_1=j_2$. In turn the relative
splitting of the collective states from the continuum is enhanced by
frustration.

\begin{figure}[t]
\begin{centering}
\includegraphics[width=0.85\columnwidth]{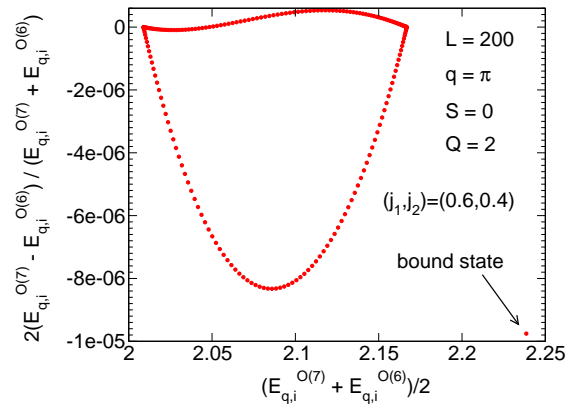}
\par\end{centering}
\caption{\label{fig9} Relative error in all two-particle energies for an FFST of
200 plaquettes at fixed total momentum $q$ for switching from $O(6)$ to $O(7)$
SE. (The apparent 'two-curve' structure is due to an oscillation of the error.)}
\end{figure}

To conclude this section we provide some measure of convergence of
the SE for the two-particle continuum in Fig. \ref{fig9}. This figure
shows the relative change in energy of all two-particle states at
a fixed total momentum versus their energy when switching from an
$O(6)$ to an $O(7)$ evaluation of the matrix elements of $H_{1,2}$.
For this we confine ourselves to $S=0$. As is obvious, the changes
are completely negligible and fully justify the use of $O(6)$ SE
for the two-particle states. We note in passing, that $O(7)$ SE will
not only lead to longer-range irreducible one-particle hoppings, but
also increase the range of the irreducible two-particle interactions
in Fig. \ref{fig7}, retaining however the rapid decay with distance
depicted there.

\section{Conclusions\label{V}}

To summarize, we have studied the strong rung-coupling regime of the frustrated
four-spin tube. Remarkably, this spin model displays several twists with respect
to conventional two-leg spin ladders. It allows for frustration induced
quantum-phase transitions into phases other than the plaquette singlet
phase. The nature of these phases remains to be explored. The structure of the
excitations is both richer, and different as compared to two-leg ladders,
including a reordering of two-particle (anti)bound states and additional
elementary excitation modes, some of which can be made to decay by reducing the
symmetry of the FFST. Finally geometric frustration on the FFST introduces a
control parameter allowing for almost localization of excitations and a
flattening of the energy landscape. As compared to potential material
realizations of the FFST in Cu$_{2}$Cl$_{4}$$\cdot$D$_{8}$C$_{4}$SO$_{2}$,
several extensions have to be considered in the future. These comprise
e.g. reducing the $C_4$-symmetry and considering the weak rung-coupling limit.

\section{Acknowledgments}

We thank D. C. Cabra, A. V. Chubukov, and O. Starykh for helpful discussions.
M. A. has been supported by CONICET (Coop. Int. R.2049/09 and PIP 1691) and by ANPCyT (PICT 1426).
W. B. thanks DFG for financial support through grant number 444 ARG-113/10/0-1.
Part of this work has been performed at the Kavli Institute for Theoretical Physics and within the
Advanced Study Group at the Max Planck Institute for the Physics of Complex
Systems, which we would like to thank for their hospitalities. The research at
KITP was supported in part by the National Science Foundation under Grant
No. NSF PHY05-51164.

\section{Effective Hamiltonian Matrix in the $Q=2$ Sector\label{app1}}

In this appendix we list some of the details necessary to evaluate the
two-particle matrix elements by SE. First, we refer to the parity of the states
in eqn. (\ref{eq:10}). Because of this we may confine ourselves to $d>0$ ($d=0$,
i.e. $|r,r\rangle^{Sm}$ is forbidden). The matrix elements of the
one(two)-particle irreducible Hamiltonian $H_{1(2)}$ are obtained by subtracting
the one(two)-particle reducible contributions from the matrix elements of
$H_{eff}$ \cite{Zheng2001a,Knetter2004b}
\begin{eqnarray}
t_{0,n} & = & t_{n,0}=\langle n|H_{1}|0\rangle=\langle n|H_{eff}-H_{0}|0\rangle\label{eq:a1}\\
 &  & a_{0,n}^{cl}-\delta_{0,n}E_{0}^{cl}\nonumber \end{eqnarray}
\begin{eqnarray}
t_{d|n,d'} & = & \langle n,n+d'|H_{2}|0,d\rangle\label{eq:a2}\\
 & = & \langle n,n+d'|H_{eff}-H_{1}-H_{0}|0,d\rangle\nonumber \\
 & = & a_{d|n,d'}^{cl}-\delta_{0,n}\delta_{d,d'}E_{0}^{cl}
-\delta_{d,n+d'}t_{0,n}^{cl}-\nonumber \\
 &  & \delta_{0,n}t_{d,n+d'}^{cl}-P\delta_{d,n}t_{0,n+d'}^{cl}-P\delta_{0,n+d'}t_{d,n}^{cl}\nonumber \end{eqnarray}
Here $a_{0,n}^{cl}=\langle n|H_{eff}|0\rangle$, $a_{d|n,d'}^{cl}=\langle n,n+d'|H_{eff}|0,d\rangle$
and $cl$ refers to the cluster on which these matrix elements are
evaluated. For $t_{0,n}$ and $t_{d|n,d'}$ to be size consistent,
the cluster has to be the largest linked cluster for a given one-
and two-particle state in real space at a given order $N$ of the
SE. The $t_{0,n}^{cl}$ in eqn. (\ref{eq:a2}) refers to $t_{0,n}$
evaluated on the \emph{same }linked cluster '$cl$' as the $a_{d|n,d'}^{cl}$
corresponding to a given $t_{d|n,d'}$ and $E_{0}^{cl}=\langle0|H_{0}|0\rangle$
is the ground state energy of that cluster. The two-particle spectrum
results from the eigensystem of\begin{equation}
^{S}\langle q,d'|H_{eff}-H_{0}|q,d\rangle^{S}={}^{S}\langle q,d'|H_{2}+H_{1}|q,d\rangle^{S}\label{eq:a3}\end{equation}
which, due to translational and spin-rotational invariance is diagonal
in $q$,$S$, and $m$ and is independent of $m$. For each $(q,S,m)$
the two matrices on the left of eqn. (\ref{eq:a3}) are hermitian
with, matrix-indices $(d',d)$. Thus, for the remainder we may consider
only $d'\leqslant d$. We begin with $h_{2d'd}={}^{S}\langle q,d'|H_{2}|q,d\rangle^{S}$.
Because a linked cluster can have at most $N+1$ sites, $h_{2d'd}=0$
for $d',d>N$. I.e. $h_{2d'd}$ is an $N\times N$ matrix. The action
of $H_{2}$ is \begin{eqnarray}
\lefteqn{H_{2}|q,d\rangle^{S}=\frac{1}{\sqrt{L}}\sum_{r}e^{iq(r+d/2)}\times}\nonumber \\
 &  & \sum_{\max(n+d',d-n)\leqslant N}t_{d|n,d'}|r+n,r+n+d'\rangle^{S}\nonumber\\
 & = & \sum_{\max(n+d',d-n)\leqslant N}t_{d|n,d'}e^{iq((d-d')/2-n)}|q,d'\rangle^{S}\label{eq:a4}\end{eqnarray}
here $n$ must be restricted to $\max(n+d',d-n)\leqslant N$, because
of the sites $(0,d,n,n+d')$ to reside within a linked cluster at
order $N$. I.e. $n$ is confined to the interval $d-N\leqslant n\leqslant N-d'$,
which has its midpoint at $n_{m}=(d-d')/2$
\begin{eqnarray}
\lefteqn{H_{2}|q,d\rangle^{S}=\{\sum_{{{\scriptstyle \max(n+d',d-n)\leqslant N}\atop
{\scriptstyle n=(d-d')/2\in\mathbb{Z}}}}t_{d|(d-d')/2,d'}+}\label{eq:a5}\\
 &  & \sum_{{{\scriptstyle \max(n+d',d-n)\leqslant N}\atop
{\scriptstyle n>(d-d')/2\in\mathbb{Z}}}}2t_{d|n,d'}\cos[q(\frac{d-d'}{2}-n)]\}
|q,d'\rangle^{S}\nonumber
\end{eqnarray}
where $t_{d|n,d'}=t_{d|d-d'-n,d'}$ has been used, which refers to
reflection symmetry of the two-particle matrix element about the midpoint,
and the addends with matrix elements $t_{d|(d-d')/2,d'}$ occurs only
if $n_{m}=(d-d')/2\in\mathbb{Z}$.

Next, we apply $H_{1}$ to a two-particle state. This will only shift
one of the particles, i.e.\begin{eqnarray}
\lefteqn{H_{1}|q,d\rangle^{S}=\sum_{{{\scriptstyle -N\leqslant n\leqslant N}\atop {\scriptstyle n\neq d}}}t_{0,n}^{cl}(e^{-iqn/2}+e^{iqn/2})}\nonumber \\
 &  & \times\frac{1}{\sqrt{L}}\sum_{r}e^{iq(r+(d-n)/2)}|r,r+d-n\rangle^{S}\nonumber \\
 & = & \sum_{{{\scriptstyle -N\leqslant n\leqslant N}\atop {\scriptstyle
n\neq d}}}2t_{0,n}^{cl}\cos(\frac{nq}{2})\mathsf{sgn}(d-n)^{S}|q,|d-n|
\rangle^{S}\nonumber \\
 \label{eq:a6}\end{eqnarray}
For each $q$, $H_{1}$ connects states $d$ and $|d-n|,\,\forall-N\leqslant n\leqslant N$.
This is equivalent to a band-matrix of width $2N+1$. For $d\geqslant N$,
the non-zero content of the columns of this matrix is independent
of $d$, for $1\leqslant d<N$ this is not so. The latter relates
to the exclusion of on-site double-occupation.

\begin{figure}[t]
\begin{centering}
\includegraphics[width=0.75\columnwidth]{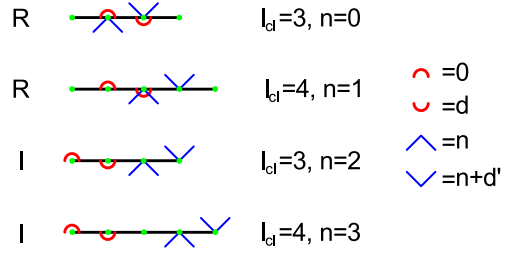}
\par\end{centering}
\caption{\label{fig10}Linked clusters with size $l_{cl}$ and with initial
(half circles) and final (wedges) two-particle states for all addend
in eqn. (\ref{eq:a2}) at order $N=4$ and for $d=d'=1$. (I)R labels
(ir)reducible graphs.}
\end{figure}

For reference we explicitly display one specific low-order matrix element
$^{S}\langle q,d'|H_{1}+H_{2}|q,d\rangle^{S}$ from eqns.  (\ref{eq:a3},
\ref{eq:a5}, \ref{eq:a6}), namely $d'=d=1$, $N=4$ and $S=0$. Figure \ref{fig10}
relates to the two-particle irreducible matrix element from $H_{2}$, and depicts
all $t_{d|n,d'}$ on their respective clusters including a label for reducing
contributions to be evaluated according to eqns. (\ref{eq:a1},
\ref{eq:a2}). These are $E_{0}^{cl},t_{0,0}^{cl},t_{1,1}^{cl}$ for $n=0$ and
$t_{0,2}^{cl}$ for $n=1$ for this particular matrix element. Note that on finite
clusters $t_{0,0}^{cl}$ and $t_{1,1}^{cl}$ are not necessarily identical.  The
one-particle irreducible matrix element from $H_{1}$ is straightforward. The
complete matrix element reads
\begin{eqnarray}
\lefteqn{^{0}\langle q,1|H_{1}+H_{2}|q,1\rangle^{0}=2+\frac{j_{1}}{6}+\frac{431j_{1}^{2}}{432}+\frac{5827j_{1}^{3}}{15552}}\nonumber \\
 &  & -\frac{18511597j_{1}^{4}}{156764160}+\frac{j_{2}}{6}-\frac{461j_{1}j_{2}}{216}-\frac{5725j_{1}^{2}j_{2}}{15552}\nonumber \\
 &  & +\frac{6030787j_{1}^{3}j_{2}}{5598720}+\frac{431j_{2}^{2}}{432}-\frac{5725j_{1}j_{2}^{2}}{15552}\nonumber \\
 &  & -\frac{51190589j_{1}^{2}j_{2}^{2}}{26127360}+\frac{5827j_{2}^{3}}{15552}+\frac{6030787j_{1}j_{2}^{3}}{5598720}\nonumber \\
 &  & -\frac{18511597j_{2}^{4}}{156764160}+\left(-\frac{14j_{1}^{2}}{27}-\frac{68j_{1}^{3}}{243}+\frac{72187j_{1}^{4}}{349920}\right.\nonumber \\
 &  & +\frac{28j_{1}j_{2}}{27}+\frac{68j_{1}^{2}j_{2}}{243}-\frac{23057j_{1}^{3}j_{2}}{17496}-\frac{14j_{2}^{2}}{27}+\frac{68j_{1}j_{2}^{2}}{243}\nonumber \\
 &  & \left.+\frac{21559j_{1}^{2}j_{2}^{2}}{9720}-\frac{68j_{2}^{3}}{243}-\frac{23057j_{1}j_{2}^{3}}{17496}+\frac{72187j_{2}^{4}}{349920}\right)\cos(q)\nonumber \\
 &  & +\left(\frac{76j_{1}^{4}}{729}-\frac{493j_{1}^{3}j_{2}}{729}+\frac{278j_{1}^{2}j_{2}^{2}}{243}-\frac{493j_{1}j_{2}^{3}}{729}\right.\nonumber \\
 &  & \left.+\frac{76j_{2}^{4}}{729}\right)\cos(2q)\label{eq:a7}\end{eqnarray}

Finally, in the $Q=2$ sector the single-particle states
$|t_{1i}^{m}\rangle,\,|t_{2i}^{m}\rangle$, and $|s_{1i}\rangle$ need to be
considered. Any one-particle irreducible Hamiltonian matrix element between the
two- and one-particle states for $Q=2$ is zero due to
orthogonality. Two-particle irreducible contributions require matrix elements of
the form\begin{equation} ^{Sm}\langle
q,d'|H_{eff}|q\rangle^{Sm}\,\,,\label{eq:a8}\end{equation} where
$|q\rangle^{Sm}=\sum_{r}e^{iqr}|r\rangle^{Sm}/\sqrt{L}$ is a one-particle state
of spin quantum numbers $S,m$ with $Q=2$. This implies a \emph{decay} of $Q=2$
single-particle states into $Q=2$ two-particle states. \emph{Up to the order
that we have performed the SE, we have not observed such decay.} In turn the
Hamiltonian matrix in the $Q=2$ sector of the four-tube is already diagonal
w.r.t. the particle number.

\bibliographystyle{apsrev}

\begin{thebibliography}{32}
\bibitem[{\citenamefont{This expressions are too long to be
  printed~explicitely. Upon~request they will be made available
electronically.}()}]{ElecAvail}
\bibinfo{author}{\bibnamefont{This
  expressions are too long to be printed~explicitely. Upon~request
they will be made available electronically}}.
\end{thebibliography}

\end{document}